\documentclass{article}
\usepackage[preprint]{corl_2022} 
\usepackage[dvipsnames]{xcolor} 
\usepackage[utf8]{inputenc} 
\usepackage[T1]{fontenc}    
\usepackage{microtype}
\usepackage{graphicx}  
\usepackage{booktabs}
\usepackage{hyperref}
\usepackage{algorithm}
\usepackage{algcompatible}
\usepackage{amsmath}
\usepackage{soul}
\usepackage[authormarkuptext=name,addedmarkup=bf,authormarkupposition=left]{changes}
\usepackage{xspace}
\usepackage{dsfont}
\usepackage{bm}
\usepackage{bbm}
\usepackage{nicefrac}       
\usepackage[scaled=.8]{beramono} 
\usepackage{amsthm}
\usepackage{amsfonts}
\usepackage{amssymb}
\usepackage{array}
\usepackage{mathtools}
\usepackage{caption}
\usepackage{subcaption}
\usepackage{fontawesome}
\usepackage{thmtools}
\usepackage{thm-restate}
\usepackage{lipsum}
\usepackage{pifont}
\usepackage{makecell}
\usepackage{tablefootnote}
\usepackage[tight-spacing=true]{siunitx}
\usepackage{multirow}
\usepackage{afterpage}
\usepackage{tabularx} 
\usepackage{url}
\usepackage{tikz}
\usepackage{indentfirst}
\usepackage{cleveref}
\usepackage{enumitem}
\usepackage{epsdice}
\usepackage{wrapfig}
\usepackage[export]{adjustbox}
\usepackage{xparse}

\makeatletter
\newcommand{\subalign}[1]{%
  \vcenter{%
    \Let@ \restore@math@cr \default@tag
    \baselineskip\fontdimen10 \scriptfont\tw@
    \advance\baselineskip\fontdimen12 \scriptfont\tw@
    \lineskip\thr@@\fontdimen8 \scriptfont\thr@@
    \lineskiplimit\lineskip
    \ialign{\hfil$\m@th\scriptstyle##$&$\m@th\scriptstyle{}##$\hfil\crcr
      #1\crcr
    }%
  }%
}
\makeatother

\Crefname{figure}{Figure}{Figures}
\crefname{figure}{Figure}{Figures}
\crefname{table}{Table}{Tables}
\crefformat{table}{Table~#1}
\crefformat{equation}{equation (#1)}
\Crefformat{Equation}{Equation (#1)}
\crefformat{algorithm}{Algorithm~#1}

\hypersetup{
  colorlinks=true,
  citecolor=Blue,
  linkcolor=Blue,
  urlcolor=Blue
}

\newcommand\sdots{\hbox to 1em{.\hss.\hss.}} 
\DeclareMathAlphabet\mathbfcal{OMS}{cmsy}{b}{n} 
\newcommand{\smallsum}{\textstyle\sum\limits}
\DeclareMathSymbol{\shortminus}{\mathbin}{AMSa}{"39}

\setitemize[0]{leftmargin=*}

\newif\ifcomments
\commentsfalse 

\ifcomments
	\newcommand{\mr}[1]{\color{orange}MR: (#1)\color{black}\xspace}  
	\newcommand{\ml}[1]{\color{red}ML: (#1)\color{black}\xspace}  
	\newcommand{\me}[1]{\color{red}ME: (#1)\color{black}\xspace}  
	\newcommand{\dk}[1]{\color{red}DK: (#1)\color{black}\xspace}  
	\newcommand{\jf}[1]{\color{red}JF: (#1)\color{black}\xspace}  
	\newcommand{\jh}[1]{\color{red}JH: (#1)\color{black}\xspace}  
	\newcommand{\gt}[1]{\color{red}GT: (#1)\color{black}\xspace}  
\else
    \newcommand{\mr}[1]{}  
	\newcommand{\ml}[1]{}  
	\newcommand{\me}[1]{}  
	\newcommand{\dk}[1]{}  
	\newcommand{\jf}[1]{}  
	\newcommand{\jh}[1]{}  
	\newcommand{\gt}[1]{}  
	\newcommand{\cc}[1]{}  
\fi
\title{Game-Theoretical Perspectives on Active Equilibria: A Preferred Solution Concept over Nash Equilibria}
\author{
  Dong-Ki Kim\\
  MID-LIDS\\
  MIT-IBM Watson AI Lab\\
  \texttt{dkkim93@mit.edu}
  \And
  Matthew Riemer\\
  IBM-Research\\
  MIT-IBM Watson AI Lab\\
  Mila\\
  \texttt{mdriemer@us.ibm.com}
  \And
  Miao Liu \\
  IBM-Research\\
  MIT-IBM Watson AI Lab\\
  \texttt{miao.liu1@us.ibm.com}
  \AND
  Jakob N. Foerster \\
  University of Oxford \\
  \texttt{jakob.foerster@eng.ox.ac.uk}
  \And
  Gerald Tesauro \\
  IBM-Research\\
  MIT-IBM Watson AI Lab\\
  \texttt{gtesauro@us.ibm.com}
  \And
  Jonathan P. How \\
  MID-LIDS\\
  MIT-IBM Watson AI Lab\\
  \texttt{jhow@mit.edu}
}

\begin{document}
\maketitle

\begin{abstract}
Multiagent learning settings are inherently more difficult than single-agent learning because each agent interacts with other simultaneously learning agents in a shared environment. An effective approach in multiagent reinforcement learning is to consider the learning process of agents and influence their future policies toward desirable behaviors from each agent's perspective. Importantly, if each agent maximizes its long-term rewards by accounting for the impact of its behavior on the set of convergence policies, the resulting multiagent system reaches an active equilibrium. While this new solution concept is general such that standard solution concepts, such as a Nash equilibrium, are special cases of active equilibria, it is unclear when an active equilibrium is a preferred equilibrium over other solution concepts. In this paper, we analyze active equilibria from a game-theoretic perspective by closely studying examples where Nash equilibria are known. By directly comparing active equilibria to Nash equilibria in these examples, we find that active equilibria find more effective solutions than Nash equilibria, concluding that an active equilibrium is the desired solution for multiagent learning settings.
\end{abstract}


\section{Introduction}\label{sec:introduction}
Multiagent reinforcement learning (MARL) provides a principled solution for settings where multiple agents interact and simultaneously learn in a shared environment~\cite{Busoniu2010,sun22romax}. 
This joint learning is particularly problematic in that each agent perceives the environment as effectively non-stationary due to the changing policies of other agents, requiring an agent to adapt its behavior with respect to the non-stationary policies of others~\citep{papoudakis19nonstationarity}. 
Importantly, the non-stationarity in multiagent learning does not result from an arbitrary stochastic process, but is caused by the joint learning process that depends on other agents' policies. 
As such, an effective agent should leverage this dependency and actively influence the learning of other agents such that their future policies evolve toward desirable policies from the agent's perspective~\cite{foerster17lola,kim21metamapg,kim22further}.

Recently, the framework by \cite{kim22further} proposes to influence the limiting policies of other agents as time approaches infinity given that these non-stationary policies will converge to a stationary periodic distribution by the end of learning. 
By learning to influence this stationary distribution, \cite{kim22further} shows that its farsighted optimization can achieve higher long-term rewards than other methods that adopt a myopic perspective that only accounts for a finite number policy updates~\cite{foerster17lola,letcher2018stable,xie20lili,kim21metamapg,wang2021influencing,lu2022modelfree} and those that neglect the policy learning process~\cite{lowe17maddpg,forester17coma,iqbal19masac,omidshafiei19teach,wadhwania2019policy,kim20hmat}. 
In particular, \cite{kim22further} formalizes this more effective joint convergence as a new game-theoretic solution concept called an active equilibrium. 
An active equilibrium provides a general solution concept that includes other standard solution concepts, such as a Nash equilibrium~\cite{Nash48}, as special cases, so the lower bound performance of active equilibria is the upper bound performance of other solution concepts.
However, this generality of active equilibrium comes at a cost because we must now additionally search for optimal learning processes and not just optimal fixed policies. As such, it is unclear when agents should search for an active equilibrium instead of other solution concepts.

\paragraph{Our contribution.} In this paper, we close the gap in understanding the benefits of active equilibria over other standard solution concepts (e.g., Nash equilibria) from a game-theoretic perspective. 
Specifically, we focus on general-sum games and directly compare solutions between active and Nash equilibria across three simple games with characteristics we believe to be quite common in multiagent RL. In each game, we demonstrate that active equilibria exist that are more effective than any Nash equilibria, indicating that it is a preferable solution concept for many multiagent settings.
\section{Background}\label{sec:active-markov-game}
This section explains relevant concepts required for understanding active and Nash equilibrium. We refer to~\cite{kim22further} for more detailed explanations.

\begin{wrapfigure}[12]{r}{0.52\linewidth}
\vskip-0.2in
\centering
\includegraphics[trim={17cm 0 0 0},clip,height=3.3cm]{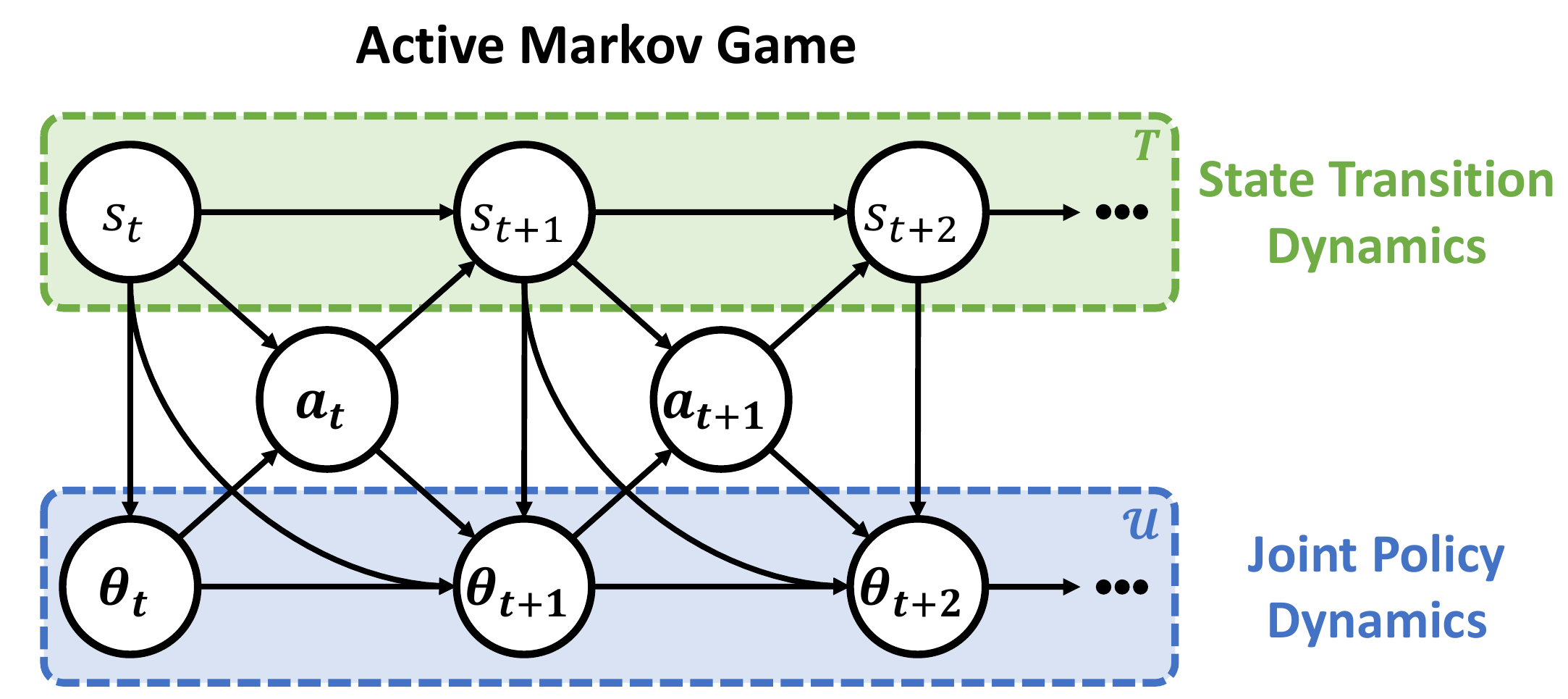}
\vskip-0.065in
\caption{Active Markov games consider that agents have non-stationary policies based on their Markovian policy update functions. Graphic credit: \cite{kim22further}}
\label{fig:active-markov-game}
\end{wrapfigure}

\paragraph{Active Markov game.} An active equilibrium provides a solution concept for an active Markov game, which represents interactions between simultaneously learning agents in MARL (see~\Cref{fig:active-markov-game}).
Formally, an $n$-agent active Markov game~\cite{kim22further} is defined as a tuple $\mathcal{M}_n\!=\!\langle\mathbfcal{I}, \mathcal{S},\mathbfcal{A},\mathcal{T},\mathbfcal{R},$ $\bm{\Theta},\mathbfcal{U}\rangle$;
$\mathbfcal{I}\!=\!\{1,\sdots,n\}$ is the set of $n$ agents;
$\mathcal{S}$ is the state space;
$\mathbfcal{A}\!=\!\times_{i \in \mathcal{I}} \mathcal{A}^{i}$ is the joint action space;
$\mathcal{T}\!:\!\mathcal{S}\!\times\!\mathbfcal{A}\!\mapsto\!\mathcal{S}$ is the state transition function;
$\mathbfcal{R}\!=\!\times_{i \in \mathcal{I}} \mathcal{R}^{i}$ is the joint reward function; 
$\bm{\Theta}\!=\!\times_{i \in \mathcal{I}} \Theta^{i}$ is the joint policy parameter space; and
$\bm{\mathcal{U}}\!=\!\times_{i \in \mathcal{I}} \mathcal{U}^{i}$ is the joint Markovian policy update function.
At each timestep $t$, each agent $i$ selects an action at a state $s_{t}\!\in\!\mathcal{S}$ according to its policy $a^i_{t}\!\sim\!\pi^{i}(\cdot|s_{t};\theta^i_{t})$ parameterized by $\theta^{i}_{t}\!\in\!\Theta^{i}$. 
Then $s_t$ transitions to $s_{t+1}$ with probability $\mathcal{T}(s_{t+1}|s_{t},\bm{a}_{\bm{t}})$, where $\bm{a}_{\bm{t}}\!=\!\{a^{i}_t,\bm{a}^{\bm{\shortminus i}}_{\bm{t}}\}$ is the joint action and the notation $\bm{\shortminus i}$ indicates all other agents except agent $i$.
Then, each agent $i$ obtains a reward according to its reward function $r^{i}_t\!=\!\mathcal{R}^i(s_t,\bm{a}_{\bm{t}})$ and updates its policy parameters according to $\mathcal{U}^{i}(\theta^{i}_{t+1}|\theta^{i}_{t},\tau^{i}_{t})$, where $\tau^{i}_{t}\!=\!\{s_{t},\bm{a}_{\bm{t}},r^{i}_{t},s_{t+1}\}$ denotes $i$'s transition. 
Note that the joint policy update function $\bm{\mathcal{U}}$ is a function of $a^{i}_{t}$, which affects the state transitions and rewards, so $i$ can actively influence future joint policies by changing its own behavior. 
Considering this active influence rather than ignoring it is the main advantage of active Markov games compared to standard Markov games~\cite{littman94markov}.

\paragraph{Stationary $k$-periodic distribution.}
The joint policy update process in an active Markov game continues until the convergence of non-stationary policies. A stationary periodic distribution over the joint space of states and policies represents the limiting policies as $t\rightarrow\infty$, where the periodic distribution can be defined as a stationary conditional distribution with respect to a period of $k$~\cite{kim22further}:
\begin{align}\label{eqn:stationary-periodic-distribution}
\begin{split}
\mu_{k}(s,\bm{\theta}|s_0,\bm{\theta_0},\ell)=p(s_t\!=\!s,\bm{\theta_t}\!=\!\bm{\theta}|s_0,\bm{\theta_{0}},\ell) \quad\forall t\!\geq\!0,s,s_0\!\in\!\mathcal{S},\bm{\theta},\bm{\theta_0}\!\in\!\bm{\Theta}.
\end{split}
\end{align}
$\ell\!=\!t\%k$ with \% denotes the modulo operation. 
$\mu_{k}(s,\bm{\theta}|s_0,\bm{\theta_0},\ell)$ satisfies the following relationship as its time averaged expectation stays stationary in the limit~\cite{kim22further}:
\begin{align}\label{eqn:stationary-periodic-distribution-property}
\begin{split}
\frac{1}{k}\smallsum_{\ell=1}^{k}&\mu_k(s_{\ell+1},\bm{\theta}_{\bm{\ell+1}}|s_0,\bm{\theta_0},\ell\!+\!1)\!=\!\frac{1}{k}\smallsum_{\ell=1}^{k}\smallsum_{s_{\ell},\bm{\theta_{\ell}}}\mu_k(s_{\ell},\bm{\theta_{\ell}}|s_0,\bm{\theta_0},\ell)\smallsum_{\bm{a_\ell}}\bm{\pi}(\bm{a_\ell}|s_{\ell};\bm{\theta_{\ell}})\\
&\quad\mathcal{T}(s_{\ell+1}|s_{\ell},\bm{a_{\ell}})\;\bm{\mathcal{U}}(\bm{\theta_{\ell+1}}|\bm{\theta_{\ell}},\bm{\tau_{\ell}})\quad\forall s_{\ell+1} \!\in\!\mathcal{S},\bm{\theta_{\ell+1}}\!\in\!\bm{\Theta}.
\end{split}
\end{align}
Note that a stationary $k$-periodic distribution provides a flexible representation for defining the limiting distribution by being able to generalize from fully stationary fixed-point convergence (i.e., $k\!=\!1$) to fully non-stationary convergence (i.e., $k\!\rightarrow\!\infty$).

\paragraph{Active equilibrium.}  
Once the joint policy converges to the stationary $k$-periodic distribution, rewards collected by this recurrent set of policies govern each agent $i$'s average reward as $t\!\rightarrow\!\infty$ \cite{kim22further}. 
Thus, an agent $i$ has an objective to find policy parameters $\theta^{i}$ and update function $\mathcal{U}^{i}$ that maximizes its expected average reward $\rho^{i}\!\in\!\mathbb{R}$ by influencing the periodic distribution at the end of learning:
\begin{align}\label{eqn:average-reward-objective}
\begin{split}
\max_{\theta^{i},\,\mathcal{U}^{i}}\rho^{i}(s,\bm{\theta},\bm{\mathcal{U}})\!:&=\!\max_{\theta^{i},\,\mathcal{U}^{i}}\lim_{T\rightarrow\infty}\!\mathbb{E}\Big[\frac{1}{T}\!\smallsum_{t=0}^{T}\!\mathcal{R}^{i}(s_t,\bm{a}_{\bm{t}})\Big|\substack{s_{0}=s,\;\bm{\theta}_{\bm{0}}=\bm{\theta},\\\bm{a_{0:T}}\sim\bm{\pi}(\bm{\cdot}|s_{0:T};\bm{\theta}_{\bm{0:T}}),\\s_{t+1}\sim\mathcal{T}(\cdot|s_{t},\bm{a_t}),\bm{\theta_{t+1}}\sim\bm{\mathcal{U}}(\cdot|\bm{\theta_{t}},\bm{\tau_{t}})}\Big]\\
&=\!\max_{\theta^{i},\,\mathcal{U}^{i}}\frac{1}{k}\smallsum_{\ell=1}^{k}\smallsum_{s_{\ell},\bm{\theta_{\ell}}}\!\!\mu_k(s_{\ell},\bm{\theta_{\ell}}|s,\bm{\theta},\ell)\smallsum_{\bm{a_\ell}}\bm{\pi}(\bm{a_\ell}|s_{\ell};\bm{\theta_{\ell}})\mathcal{R}^{i}(s_{\ell},\bm{a_{\ell}}),
\end{split}
\end{align}
where $T$ denotes the time horizon. 
If all agents maximize the active average reward objective in \Cref{eqn:average-reward-objective}, agents arrive at an \textit{active equilibrium} \cite{kim22further}, associated with joint policy parameters $\bm{\theta^{*}}\!=\!\{\theta^{i*},\bm{\theta^{\shortminus i*}}\}$ and update function $\bm{\mathcal{U}^{*}}\!=\!\{\mathcal{U}^{i*},\bm{\mathcal{U}^{\shortminus i*}}\}$, where no agents can further optimize its active average reward:
\begin{align}
\begin{split}
\rho^{i}(s,\theta^{i*},\bm{\theta}^{\bm{\shortminus i*}},\mathcal{U}^{i*},\bm{\mathcal{U}^{\shortminus i*}})\!\geq\!\rho^{i}(s,\theta^{i},\bm{\theta}^{\bm{\shortminus i*}},\mathcal{U}^{i},\bm{\mathcal{U}^{\shortminus i*}}) \quad\forall i\!\in\!\mathcal{I},s\!\in\!\mathcal{S},\theta^{i}\!\in\!\Theta^{i},\mathcal{U}^{i}\!\in\!\mathbb{U}^{i}.
\end{split}
\end{align}
where $\mathbb{U}^{i}$ denotes the space of agent $i$'s update functions. 
An active equilibrium generalizes from a stationary convergence (for $k=1$) to a non-stationary convergence (for $k>1$). 
We note that an active equilibrium is related to the framework by~\cite{lu2022modelfree} that also considers the active influence concept in a meta-learning setting. 
\cite{lu2022modelfree} models finite joint policy updates in learning the meta-parameters and if this modeling is performed over infinite policy updates, the resulting meta-parameters would find a similar solution to an active equilibrium.

\paragraph{Nash equilibrium.} A Nash equilibrium~\cite{Nash48} corresponds to a special case of an active equilibrium when $k=1$ and the joint update function $\bm{\mathcal{U}}$ is the identity function (i.e., a stationary equilibrium).
Specifically, an agent $i$ now has an objective with respect to stationary policies:
\begin{align}\label{eqn:average-reward-objective-nash}
\begin{split}
\max_{\theta^{i}}\rho_{\bm{\theta}}^{i}(s)\!:&=\!\max_{\theta^{i}}\lim_{T\rightarrow\infty}\!\mathbb{E}\Big[\frac{1}{T}\!\smallsum_{t=0}^{T}\!\mathcal{R}^{i}(s_t,\bm{a}_{\bm{t}})\Big|\substack{s_{0}=s,\\\bm{a_{0:T}}\sim\bm{\pi}(\bm{\cdot}|s_{0:T};\bm{\theta}),\\s_{t+1}\sim\mathcal{T}(\cdot|s_{t},\bm{a_t})}\Big]\\
&=\!\max_{\theta^{i}}\smallsum_{s^{\prime}}\!\!\mu_{k=1}(s^{\prime},\bm{\theta}|s,\bm{\theta},\ell)\smallsum_{\bm{a}}\bm{\pi}(\bm{a}|s;\bm{\theta})\mathcal{R}^{i}(s,\bm{a}),
\end{split}
\end{align}
where the subscript $\bm{\theta}$ denotes the implicit dependence on stationary policy parameters. 
Then, if all agents maximize the stationary average reward objective in~\Cref{eqn:average-reward-objective-nash}, agents arrive at a \textit{Nash equilibrium}, associated with joint policy parameters $\bm{\theta^{*}}\!=\!\{\theta^{i*},\bm{\theta^{\shortminus i*}}\}$, where no agents can further improve its stationary average reward:
\begin{align}
\begin{split}
\rho^{i}_{\theta^{i*},\bm{\theta}^{\bm{\shortminus i*}}}(s)\!\geq\!\rho^{i}_{\theta^{i},\bm{\theta}^{\bm{\shortminus i*}}}(s) \quad\forall i\!\in\!\mathcal{I},s\!\in\!\mathcal{S},\theta^{i}\!\in\!\Theta^{i}.
\end{split}
\end{align}
\section{Understanding the Benefits of Active Equilibria}\label{sec:method}
We study whether the active equilibrium is a more desired solution concept than the Nash equilibrium. To answer this question, we focus on the following general-sum game settings and directly compare active and Nash equilibria to show the beneficial characteristics of active equilibria:
\begin{itemize}[leftmargin=*, wide, labelindent=0pt, topsep=0pt]
    \itemsep 0in 
    \item A finite repeated game of the prisoner's dilemma (see \Cref{fig:ipd-domain})
    \item A finite repeated game of Bach or Stravinsky (see \Cref{fig:ibs-domain})
    \item A periodic game with a period of $2$ (see \Cref{fig:periodic-odd-domain,fig:periodic-even-domain})
\end{itemize}

\textbf{Remark 1 (Social dilemma game).} \textit{Active equilibria include the tit-for-tat strategy, enabling mutual cooperation in a finitely repeated game of prisoner's dilemma.}

Consider playing a finitely repeated game of prisoner’s dilemma (see \Cref{fig:ipd-domain}). 
We do not study the infinitely repeated prisoner's dilemma because the folk theorem states that there are infinitely many Nash equilibria in the infinite horizon settings~\cite{gametheorybook}, making it impossible to compare active and Nash equilibria. 

Given a finite horizon, the only Nash equilibrium is the mutual defection. This result can be checked using backward induction \cite{gametheorybook}, where Nash achieves the undesirable result as the equilibrium learns stationary policies assuming current policies cannot affect future policies.
By contrast, an active equilibrium corresponds to $\bm{a_0}=\{C,C\}$ associated with the following update function: $\mathcal{U}^{i}(\theta^{i}_{t+1}|\theta^{i}_{t},\{C,C\})\!=\!C$, $\mathcal{U}^{i}(\theta^{i}_{t+1}|\theta^{i}_{t},\{C,D\})\!=\!D$, $\mathcal{U}^{i}(\theta^{i}_{t+1}|\theta^{i}_{t},\{D,C\})\!=\!C$, $\mathcal{U}^{i}(\theta^{i}_{t+1}|\theta^{i}_{t},\{D,D\})\!=\!D$. 
This equilibrium achieves an active equilibrium with no incentive to deviate from $\bm{\theta_0}$ and $\bm{\mathcal{U}}$.
We observe that active equilibria include a tit-for-tat strategy, where the deviation penalty is embedded in $\bm{\mathcal{U}}$, achieving the desired result of mutual cooperation. 
We also note that this mutual cooperation convergence coincides with the cooperation result in LOLA~\cite{foerster17lola} because both the active equilibrium and LOLA consider the active influence.

\textbf{Remark 2 (Multiple equilibria game).} \textit{Active equilibria include more efficient solutions than Nash.}

Consider playing a finitely repeated game of Bach or Stravinsky (see \Cref{fig:ibs-domain}). 
Given a finite horizon, the pure Nash equilibria are: $\bm{a_t}=\{B,B\}$ $\forall t$ or $\bm{a_t}=\{S,S\}$ $\forall t$. However, these solutions are unfair because only one agent consistently receives higher rewards than the other. There is also the mixed Nash equilibrium: agent $i$ acts $B$ and $S$ actions with $2/3$ and $1/3$ probabilities, respectively, and agent $\shortminus i$ acts $B$ and $S$ actions with $1/3$ and $2/3$ probabilities, respectively, $\forall t$. However, this solution is inefficient as the average rewards for both agents are $(2/3,2/3)$, receiving lower rewards than the rewards achieved by the pure Nash solutions. 
By contrast, an active equilibrium in this example corresponds to $\bm{a_0}=\{B,B\}$ associated with the update function: $\mathcal{U}^{i}(\theta^{i}_{t+1}|\theta^{i}_{t},\{B,B\})\!=\!S$, $\mathcal{U}^{i}(\theta^{i}_{t+1}|\theta^{i}_{t},\{B,S\})\!=\!S$, $\mathcal{U}^{i}(\theta^{i}_{t+1}|\theta^{i}_{t},\{S,B\})\!=\!B$, $\mathcal{U}^{i}(\theta^{i}_{t+1}|\theta^{i}_{t},\{S,S\})\!=\!B$. 
Both agents receive rewards of $(3/2,3/2)$ at this active equilibrium, achieving a more fair and efficient Nash equilibrium. 

\begin{figure}[t!]
\captionsetup[subfigure]{skip=0pt, aboveskip=13pt}
    \begin{subfigure}[b]{0.24\linewidth}
        \centering
        \setlength{\tabcolsep}{2pt}
        \begin{tabular}[b]{cc|cc}
        \multicolumn{2}{c}{} & \multicolumn{2}{c}{}\\
        \parbox[t]{2mm}{\multirow{3}{*}{}} 
            &       & $C$       & $D$       \\\cline{2-4}
        \rule{0pt}{10pt}    & $C$   & $(\shortminus 1,\shortminus 1)$ & $(\shortminus 3,0)$  \\
            & $D$   & $(0,\shortminus 3)$  & $(\shortminus 2,\shortminus 2)$ 
        \end{tabular}
        \vskip-0.13in
        \caption{Prisoner's dilemma}
        \label{fig:ipd-domain}
    \end{subfigure}
    \begin{subfigure}[b]{0.24\linewidth}
        \centering
        \setlength{\tabcolsep}{2pt}
        \begin{tabular}[b]{cc|cc}
        \multicolumn{2}{c}{} & \multicolumn{2}{c}{}\\
        \parbox[t]{2mm}{\multirow{3}{*}{}} 
            &       & $B$       & $S$       \\\cline{2-4}
        \rule{0pt}{10pt}    & $B$   & $(2,1)$ & $(0,0)$  \\
            & $S$   & $(0,0)$  & $(1,2)$ 
        \end{tabular}
        \vskip-0.13in
        \caption{Bach or Stravinsky}
        \label{fig:ibs-domain}
    \end{subfigure}
        \begin{subfigure}[b]{0.24\linewidth}
            \centering
            \setlength{\tabcolsep}{2pt}
            \begin{tabular}[b]{cc|cc}
            \multicolumn{2}{c}{} & \multicolumn{2}{c}{}\\
            \parbox[t]{2mm}{\multirow{3}{*}{}} 
                &       & $A$       & $B$       \\\cline{2-4}
            \rule{0pt}{10pt}    & $A$   & $(2,2)$ & $(0,0)$  \\
                & $B$   & $(0,0)$  & $(1,1)$ 
            \end{tabular}
            \vskip-0.13in
            \caption{Periodic game: odd $t$}
            \label{fig:periodic-odd-domain}
        \end{subfigure}
        \begin{subfigure}[b]{0.24\linewidth}
            \centering
            \setlength{\tabcolsep}{2pt}
            \begin{tabular}[b]{cc|cc}
            \multicolumn{2}{c}{} & \multicolumn{2}{c}{}\\
            \parbox[t]{2mm}{\multirow{3}{*}{\rotatebox[origin=c]{90}{}}} 
                &       & $A$       & $B$       \\\cline{2-4}
            \rule{0pt}{10pt}    & $A$   & $(1,1)$ & $(0,0)$  \\
                & $B$   & $(0,0)$  & $(2,2)$ 
            \end{tabular}
            \vskip-0.13in
            \caption{Periodic game: even $t$}
            \label{fig:periodic-even-domain}
        \end{subfigure}
    \caption{Payoff tables for (a) prisoner's dilemma, (b) Bach or Stravinsky, (c) and (d) periodic game.}
    \vskip-0.15in
\end{figure}

\textbf{Remark 3 (Periodic game).} \textit{Active equilibria have an inherent benefit over fixed policy Nash equilibria when the game itself is non-stationary or periodic in nature.}

Periodic games model realistic settings where the game itself is not fixed and has a cyclic payoff matrix $A$ with a finite period $k$ (i.e., $A(t)$ = $A(t + k)$) \cite{fiez21periodic}. 
Specifically, we consider a periodic game with a period of $2$ alternating between \Cref{fig:periodic-odd-domain} and \Cref{fig:periodic-even-domain} for a finite horizon. 
Nash equilibria correspond to either $\bm{a_t}=\{A,A\}$ $\forall t$ or $\bm{a_t}=\{B,B\}$ $\forall t$, achieving the sub-optimal average rewards of $(2/3,2/3)$ for both agents.
By contrast, an active equilibrium corresponds to $\bm{a_0}=\{A,A\}$ with the update function: $\mathcal{U}^{i}(\theta^{i}_{t+1}|\theta^{i}_{t},\{A,A\})\!=\!B$, $\mathcal{U}^{i}(\theta^{i}_{t+1}|\theta^{i}_{t},\{A,B\})\!=\!B$, $\mathcal{U}^{i}(\theta^{i}_{t+1}|\theta^{i}_{t},\{B,A\})\!=\!A$, $\mathcal{U}^{i}(\theta^{i}_{t+1}|\theta^{i}_{t},\{B,B\})\!=\!A$. 
At this active equilibrium, agents show a non-stationary (cyclic) convergence, achieving average rewards of $(2,2)$ for both agents.
Hence, active equilibria can learn non-stationary convergence, which has the benefit over stationary equilibrium (e.g., Nash) when the underlying environment itself has non-stationary behavior.
\section{Conclusion}
In this paper, we have studied the general solution concept of active equilibria by closely looking at a few simple example games that highlight key characteristics encountered in multiagent RL. 
These examples show if the game structure has the social dilemma property, multiple equilibria, or highlights non-stationarity, then active equilibria may exist with better performance than those that can be found within the traditional solution concept of Nash equilibria. As such, this paper contributes to closing the gap in understanding the benefits of active equilibria compared to Nash equilibria. We conclude that an active equilibrium is the desired solution for multiagent learning settings. Our future work will include developing systematic algorithms (e.g., value iteration methods assuming given models) for finding an active equilibrium in general-sum and active Markov games.

\subsubsection*{Acknowledgments}
Research funded by IBM (as part of the MIT-IBM Watson AI Lab initiative).


\bibliography{main}

\end{document}